\documentclass{iopart}  

\usepackage[english]{babel}
\usepackage{verbatim}
\usepackage[dvips]{color}
\usepackage{graphicx}

\begin{document}
\title[]{Measurement of
$^{25}\mathrm{Mg}(\mathrm{p},\gamma)^{26}\mathrm{Al}$ resonance
strengths via gamma spectrometry}
\author{A Formicola$^1$, A Best$^2$, G Imbriani$^3$, M Junker$^1$, D Bemmerer$^4$, R Bonetti$^5$, C Broggini$^6$, A Caciolli$^6$, F Confortola$^7$, P Corvisiero$^7$, H Costantini$^7$,
Z Elekes $^8$, Zs Fulop$^8$, G Gervino$^9$, A Guglielmetti$^5$, Gy
Gy\"{u}rky$^8$, C Gustavino$^1$, A Lemut$^7$, B Limata$^{10}$, M
Marta$^4$, C Mazzocchi$^5$, R Menegazzo$^6$, P Prati$^7$, V
Roca$^{10}$, C Rolfs$^2$, C Rossi Alvarez $^6$, E Somorjai$^8$, O
Straniero$^{3}$, F Strieder$^2$, F Terrasi$^{10}$, H P
Trautvetter$^2$ }

\address{$^{1}$ Laboratori Nazionali del Gran Sasso, INFN, Assergi Italy} \address{$^{ 2}$ Insitut f\"{u}r Physik mit
Ionenstrahlen, Ruhr-Universit\"{a}t-Bochum, Bochum Germany}
\address{$^{3}$ Osservatorio Astronomico di Collurania, Teramo and
INFN Sezione di Napoli, Italy}
\address{$^{4}$
Forschungszentrum Dresden-Rossendorf, Dresden, Germany}
\address{$^{5}$
Istituto di Fisica Generale Applicata, Universit\`{a} di Milano
and INFN Sezione di Milano, Italy}
\address{$^{6}$
Istituto Nazionale di Fisica Nucleare (INFN), Sezione di Padova,
Italy}
\address{$^{7}$
Universit\`{a} di Genova and INFN Sezione di Genova,Genova Italy}
\address{$^{8}$
Institue of Nuclear Research (ATOMKI),Debrecen,Hungary}
\address{$^{9}$
Dipartimento di Fisica Sperimentale, Universit\`{a} di Torino and
IFNN Sezione di Torino, Torino Italy}
\address{$^{10}$
Dipartimento di Scienze Fisiche, Universit\`{a} Napoli "Federico
II" and INFN Sezione di Napoli, Napoli, Italy}

\ead{formicola@lngs.infn.it}

\begin{abstract}
The COMPTEL instrument performed the first mapping of the 1.809
MeV photons in the Galaxy, triggering considerable interest in
determing the sources of interstellar $^{26}$Al. The predicted
$^{26}$Al is too low compared to the observation, for a better
understanding more accurate rates for the
$^{25}\mathrm{Mg}(\mathrm{p},\gamma)^{26}\mathrm{Al}$ reaction are
required.\\ The
$^{25}\mathrm{Mg}(\mathrm{p},\gamma)^{26}\mathrm{Al}$ reaction has
been investigated at the resonances at E$_{r}$\footnote{As a
general notation in this work, E$_{r}$ is the resonance energy in
the center of mass system.}$ = 745,418,374,304$ keV
at Ruhr-Universit\"{a}t-Bochum using a Tandem accelerator and a
$4\pi$ NaI detector. In addition the resonance at E$_{r} = 189$
keV has been measured deep underground laboratory at Laboratori
Nazionali del Gran Sasso, exploiting the strong suppression of
cosmic background. This low resonance has been studied with the
400 kV LUNA accelerator and a HPGe detector. The preliminary
results of the resonance strengths will be reported.
\end{abstract}

\section {The $^{25}\mathrm{Mg}(\mathrm{p},\gamma)^{26}\mathrm{Al}$ reaction}
The $^{25}\mathrm{Mg}(\mathrm{p},\gamma)^{26}\mathrm{Al}$ is the
slowest reaction of the Mg-Al cycle. The {$^{26}\mathrm{Al}$}
ground state (T$_{1/2}$ = $7*10^{5}$ yr) decays via $\beta^{+}$
and EC to the the $2^{+}$ first excited state of
{$^{26}\mathrm{Mg}$}. This level decays to the ground state of
{$^{26}\mathrm{Mg}$} with the emission of a 1.809 MeV
$\gamma$-ray, one of the most important lines in $\gamma$
astronomy.
The direct observation of this radiation from COMPTEL \cite{compt1} and INTEGRAL \cite{compt2}
instruments provides an evidence that $^{26}\mathrm{Al}$ production is still active on a large scale. Moreover, the
observation of {$^{26}\mathrm{Mg}$} isotopic enrichment (extinct
{$^{26}\mathrm{Al}$}) in carbonaceous meteorites \cite{meteo} shows that the $^{26}\mathrm{Al}$ had been produced
before the formation of the solar system  ($4.6*10^{9}$ yr). Any astrophysical scenario must be concordant with both
observations.
Stellar nucleosynthesis studies have not yet
identified which one of the possible {$^{26}\mathrm{Al}$} sources could explain the observed evidences. Existing
stellar models predict that 30-50\% of the production of $^{26}\mathrm{Al}$ comes from hydrogen-burning shell (HBS)
of massive stars (core collapse Supernovae and Wolf-Rayet stars). The remaining should come elsewhere, for example
from HBS of low mass AGB or from the nucleosynthesis of Novae.
Hence, solving the controversy for different astrophysical
production sites of {$^{26}\mathrm{Al}$} demands a better
understanding of the rates for the
$^{25}\mathrm{Mg}(\mathrm{p},\gamma)^{26}\mathrm{Al}$ reaction.
\begin{figure}[ptb]
    \centering
        \includegraphics[width=0.50\textwidth]{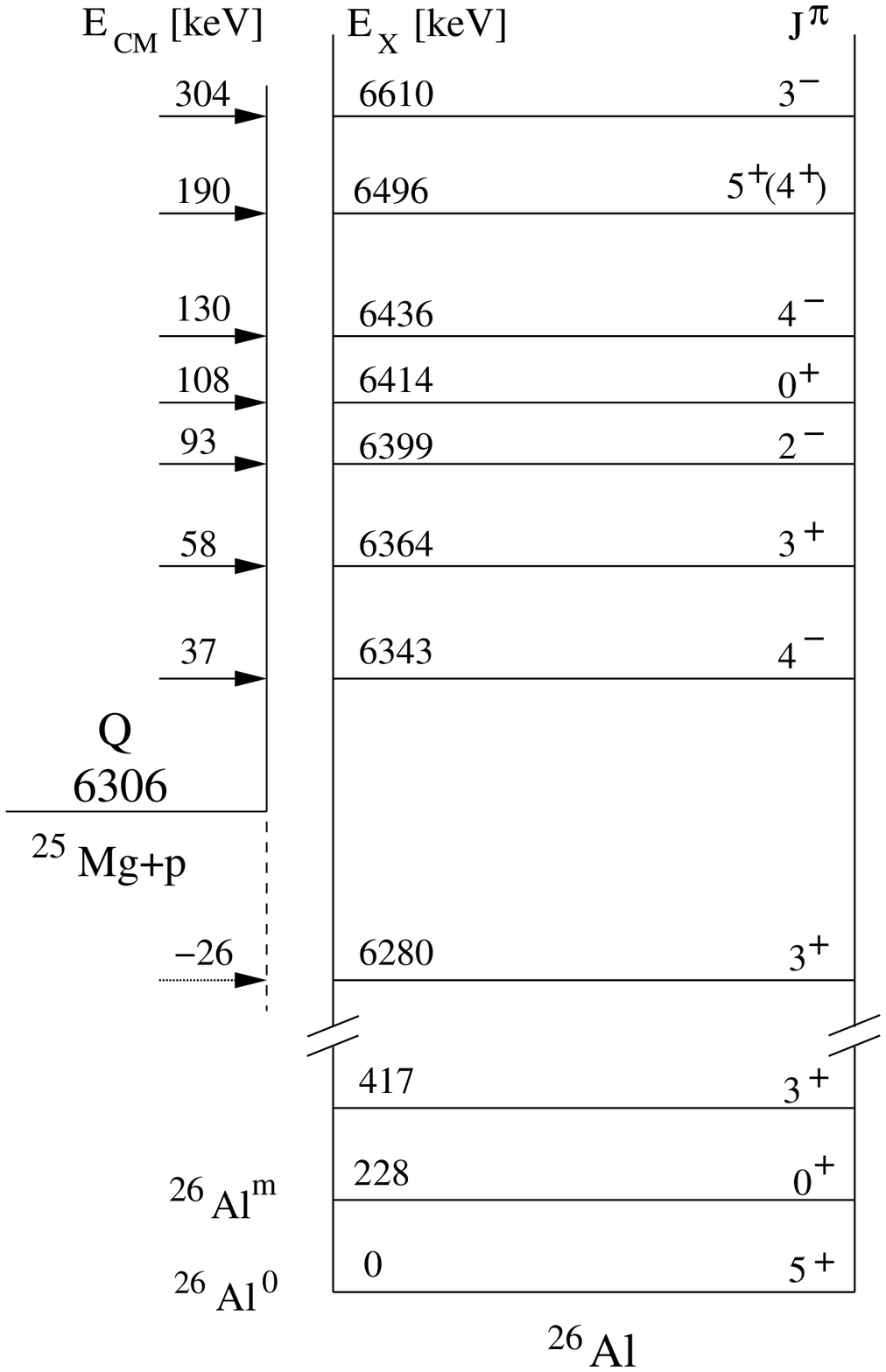}
    \caption{Level scheme of $^{26}\mathrm{Al}$.}
    \label{livelli}
\end{figure}

Figure ~\ref{livelli} shows the level scheme of
{$^{26}\mathrm{Al}$}. The Q-value of the reaction is Q = 6306 keV.
Any internal transition from the isomeric state
{$^{26}\mathrm{Al}$}$^{m}$ (T$_{1/2}$ = 6.35 s) to the ground
state of {$^{26}\mathrm{Al}$} are inhibited due to the large spin
difference. In the past the levels down to 6496 keV, corresponding
to the resonance E$_{r}$ = 189 keV have been directly and
indirectly studied \cite{cham2, endt, ili, pow}. A recent
publication by Arazi et al. \cite{ara} making use of the AMS
technique gives resonance strengths down to the resonance energy
at E$_{r}$ = 189 keV. The main difference, between AMS results and
the previous data in literature, concerns the value of the
resonance strength at E$_{r}$ = 189 keV, where Arazi et al.
\cite{ara} quoted a value about 5 times smaller than the value
published previously \cite{nacre, ili}.
\section{Measurements of $^{25}\mathrm{Mg}(p,\gamma)^{26}\mathrm{Al}$ strengths}
\begin{figure}[ptb]
    \centering
        \includegraphics[width=0.8\textwidth]{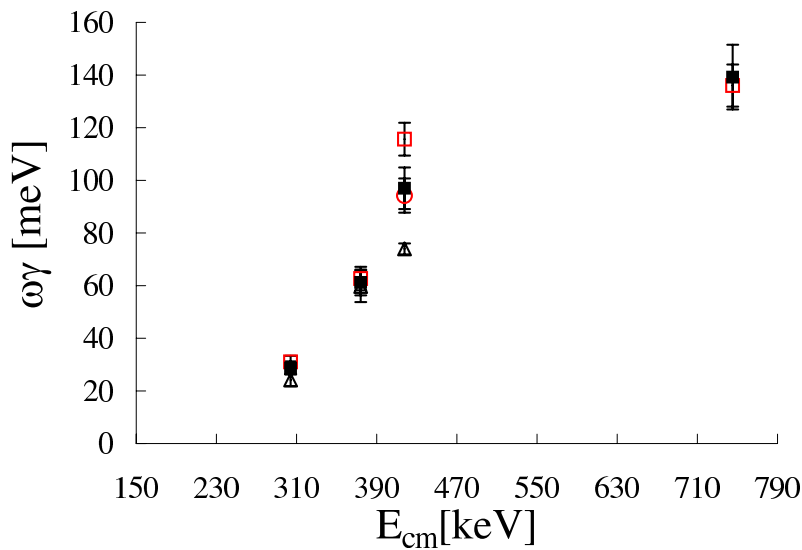}
    \caption{Resonance strengths versus the energy in the center of mass. Filled in data points are the results from
Bochum analysis \cite{andre}, the open triangles are from AMS measurements \cite{ara}, the open square are from Nacre compilation \cite{nacre},
and the open circle is the experimental value measured by Powell et al. \cite{pow}.}
\label{sum}
\end{figure}

In the framework of LUNA, the resonances at E$_{r}$ = 304, 374,
418, and 745 keV have been investigated \cite{andre} at
Ruhr-Universit\"{a}t-Bochum using the 4MV Dynamitron-Tandem
accelerator, with an average proton beam current of 100nA, and a
$12^{"}$x$12^{"}$ $4\pi$ NaI summing crystal \cite{mehr}. The Mg
targets have been produced evaporating MgO powder isotopically
enriched in $^{25}$Mg (98\%) on Ta backing. The same technique for
target production was also used for low energy measurements. The
efficiency in the energy range 3000keV$<E_{\gamma}<$7000keV was
determined to be about 70\% using a Monte Carlo code based on
Geant4 \cite{mc} simulation. In the simulation the branching
ratios of the resonances and known levels in $^{26}Al$ from
literature \cite{endt88, ili} have been used to reproduce the
complex decay scheme of the
$^{25}\mathrm{Mg}(\mathrm{p},\gamma)^{26}\mathrm{Al}$ reaction. In
this way the simulated $\gamma$-spectra could be directly fitted
to the experimental spectra and no additional information about
average multiplicity was necessary (see e.g. \cite{spyrou}).  The
uncertainty of this procedure was estimated by variing the
branching ratios in a reasonable range \cite{andre}: no
significant influence was observed. The code was tested with
calibrated sources \cite{andre} and, moreover, a perfect agreement
with the results of \cite{mehr} was obtained. A detailed
discussion of these measurements will be presented in a
forthcoming article \cite{best}.
\begin{figure}[ptb]
    \centering
        \includegraphics[width=0.8\textwidth]{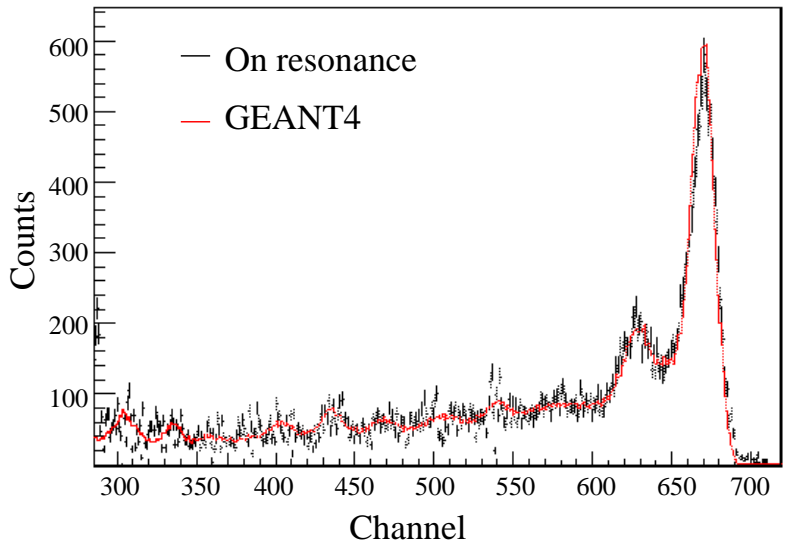}
    \caption{Comparison between experimental spectrum (black curve) acquired at E$_{r}$ = 304 keV and a simulated one
(red curve).}
    \label{317}
\end{figure}
Figure \ref{317} shows a comparison between the experimental
spectrum recorded at E$_{r}$ = 304 keV, and a simulated one. The
strength values are shown in fig. \ref{sum}.
The $\omega\gamma$ results are in good agreement with previous
work in particular \cite{pow,ili} and disagree with \cite{ara}.
\\In addition, the resonance at E$_{r}$= 189 keV has been
investigated using the 400kV LUNA accelerator facility installed
in a deep underground laboratory in the Laboratori Nazionali del
Gran Sasso. The peculiarities of this accelerator have been
described elsewhere \cite{alba}. The detector used was a HPGe
detector (119\% efficiency), placed at $\theta=55^{\circ}$
relative to the beam direction. The distance between the target
and the front face of the detector was 3.5 cm, in order to
guarantee a high detection efficiency. Due to this geometry significant corrections due to the summing effect had to
be applied.
The efficiency curve for the measurement of the E$_{r}$= 189 keV resonance energy was determined using the
$^{14}$N(p,$\gamma$)$^{15}$O (E$_{r}$= 259 keV) and $^{24}$Mg(p,$\gamma$)$^{25}$Al (E$_{r}$= 224 keV) reactions as
well as $^{137}$Cs and $^{207}$Bi calibrated sources. Spectra were collected with the detector at three different
distances from the target, with the aim of investigating in detail the summing in and the summing out effects.
In order to prevent build-up of impurities on the target, a LN-cooled trap was mounted directly to
the front face of the target. Repeated measurement of the
resonance profile during long-term high beam bombardments (average
current 250$\mu$A) allowed for monitoring the $^{25}$Mg target
quality and purity.
\begin{figure}[ptb]
    \centering
        \includegraphics[width=0.8\textwidth]{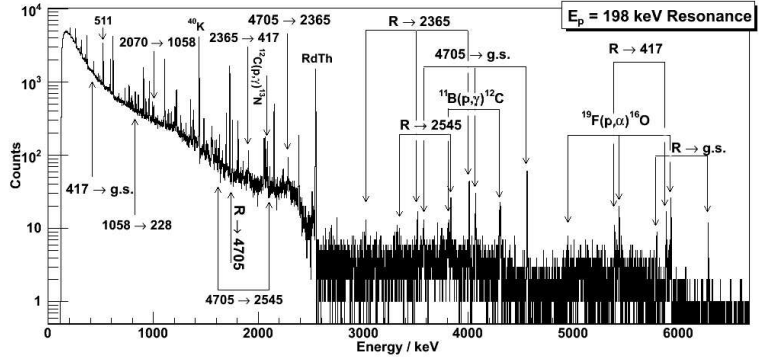}
    \caption{Undergroumd $\gamma$-spectrum of the $^{25}\mathrm{Mg}(\mathrm{p},\gamma)^{26}\mathrm{Al}$ reaction
recorded at resonance energy E$_{r}$ = 189 keV.}
    \label{spettro}
\end{figure}
Figure \ref{spettro} shows the experimental spectrum acquired at
E$_{r}$= 189 keV with a total charge of 25C. This set-up improves
the knowledge of the $\gamma$-ray cascade structure of the
resonance. We observed the complex $\gamma$-ray cascade including
the transition to the ground state. We also were able to identify
the principal background contamination reactions:
$^{11}\mathrm{B}(p,\gamma)^{12}\mathrm{C}$ and
$^{19}\mathrm{F}(p,\alpha\gamma)^{16}\mathrm{O}$. After a
preliminary analysis we are able to quote the branching ratio for
the ground state transition to be 6\%. Due to the difficulties in
the determination of summing out correction, we are able to give a
preliminary range for the resonance strength at E$_{r}$ = 189 keV,
$6.8*10^{-7}eV<\omega\gamma<7.7*10^{-7}eV$. This range is in
perfect agreement with the value given by Iliadis et al $(7.4 \pm
1.0)*10^{-7}eV$ \cite{ili}. The analysis of the full data set is
still in progress, as well as a new $\gamma$-measurement with a
BGO summing crystal at LNGS, aiming the first direct detection of
the resonance at E$_{r}$ = 93 keV.
\section{Acknowledgments}
"We would like to thank Helmut Baumeister, University of
M\"{u}nster (Germany) and Massimo Loriggiola, INFN Laboratori
Nazionali di Legnaro (Itlay) for the Magnesium target production.
This work was supported by INFN and in part by the European Union
(TARI RII3-CT-2004-506222), the Hungarian Scientific Research Fund
(K68801 and T49245) and the Deutsche Forschung Gemeinschaft (DFG)
(Ro429/41)."

-

\end{document}